\documentclass[a4paper,11pt,twoside]{book}

\usepackage[latin1]{inputenc}

\usepackage{mathptmx}

\DeclareMathAlphabet{\mathsc}{OT1}{cmr}{m}{sc}
\def\testbx{bx}%
\DeclareRobustCommand{\ion}[2]{%
\relax\ifmmode
\ifx\testbx\f@series
{\mathbf{#1\,\mathsc{#2}}}\else
{\mathrm{#1\,\mathsc{#2}}}\fi
\else\textup{#1\,{\mdseries\textsc{#2}}}%
\fi}

\usepackage{dropping}

\usepackage[a4paper,twoside,dvips,heightrounded]{geometry}

\usepackage{setspace}
\usepackage[multiple,stable]{footmisc}

\usepackage[small,bf]{caption}
\usepackage[footnotesize,bf,FIGBOTCAP]{subfigure}

\usepackage{fancyhdr}

\fancypagestyle{plain}{%
  \fancyhf{}
  
  \fancyhf[FLE,FRO]{\hfill\slshape\thepage}}

\makeatletter
\renewcommand{\chaptermark}[1]{\markboth{\@chapapp\ \thechapter.\ \ #1}{}}
\makeatother

\usepackage{sectsty}
\subsubsectionfont{\vspace{5pt}\slshape\mdseries}

\usepackage[dvips]{color}
\usepackage[grey]{quotchap}

\definecolor{chaptergrey}{rgb}{0.1,0.1,0.6}

\renewcommand{\numberalign}{\raggedleft}
\renewcommand{\textalign}{\raggedleft}

\usepackage[subfigure]{tocloft}

\tocloftpagestyle{plain}

\makeatletter
\renewcommand{\cftmarktoc}{\@mkboth{\contentsname}{\contentsname}}
\renewcommand{\cftmarklof}{\@mkboth{\listfigurename}{\listfigurename}}
\renewcommand{\cftmarklot}{\@mkboth{\listtablename}{\listtablename}}
\makeatother

\renewcommand{\contentsname}{Contents}

\renewcommand{\listfigurename}{List of Figures}

\renewcommand{\listtablename}{List of Tables}

\usepackage{tocbibind}

\makeatletter
\def\@makeschapterhead#1{%
  {\parindent \z@ \centering
    \normalfont
    \interlinepenalty\@M
    \huge \bfseries \sc #1\par\nobreak
    \vskip 10\p@
  }}
\makeatother

\usepackage[toc]{appendix}

\noappendicestocpagenum

\usepackage{amsmath}

\allowdisplaybreaks

\makeatletter
\@ifundefined{mathindent}{\relax}{\setlength{\mathindent}{0pt}}
\makeatother

\usepackage{amssymb}

\usepackage[dvips,final]{graphicx}

\usepackage{epsfig}
\usepackage{rotating}

\usepackage[round]{natbib}

\usepackage{epigraph}

\renewcommand{\epigraphsize}{\small}

\newlength{\myepigraphwidthfrontend}
\renewcommand{\textflush}{flushepinormal}

\renewcommand{\epigraphflush}{flushleft}

\renewcommand{\sourceflush}{flushright}

\usepackage{array}

\usepackage{booktabs}
\usepackage{url}

\makeatletter
\renewcommand*{\cleardoublepage}{%
  \clearpage
  \if@twoside
    \ifodd\c@page
    \else
      \thispagestyle{empty}     
      \hbox{}\newpage
      \if@twocolumn
        \hbox{}\newpage
      \fi
    \fi
  \fi
}
\makeatother

\usepackage{amssymb}
\usepackage{aas_macros}

\newcommand{\thesistitle}{Multi-dimensional analysis of the chemical and
  physical properties of spiral galaxies}

\newcommand{\thesisauthor}{Fernando Fabián Rosales Ortega}

\newcommand{\thesisdate}{November 5, 2009}

\usepackage{placeins}
\usepackage{dcolumn}
\usepackage{multirow}
\usepackage{endnotes}

\usepackage[ps2pdf,backref,pdfpagemode=UseOutlines,breaklinks=true]{hyperref}

\hypersetup{citecolor={blue},linkcolor={blue},urlcolor={blue}}     

\hypersetup{
  colorlinks=true,             
  bookmarksnumbered=true,      
  bookmarksopen=false,         
  pdftitle={\thesistitle},     
  pdfauthor={\thesisauthor},   
  pdfpagelayout=TwoPageRight,  
  pdfdisplaydoctitle,          
  pdfsubject={Ph.D. dissertation, Institute of Astronomy, University of Cambridge}
}

\usepackage[all]{hypcap}

\newcommand{\lam}{$\lambda$}

\newcommand{\hi}{H\,{\footnotesize I}~}
\newcommand{\hh}{H\,{\footnotesize II}~}
\newcommand{\hei}{He\,{\footnotesize I}~}
\newcommand{\heii}{He\,{\footnotesize II}~}

\newcommand{\oii}{[O\,{\footnotesize II}]~}
\newcommand{\nii}{[N\,{\footnotesize II}]~}
\newcommand{\sii}{[S\,{\footnotesize II}]~}
\newcommand{\oiii}{[O\,{\footnotesize III}]~}

\defcitealias{Sanchez:2007p1276}{San07}

\defcitealias{Kewley:2002p311}{KD02}
\defcitealias{McGaugh:1991p314}{M91}
\defcitealias{Zaritsky:1994p333}{Z94}
\defcitealias{Kobulnicky:2004p1700}{KK04}
\defcitealias{Denicolo:2002p361}{D02}
\defcitealias{Pettini:2004p315}{PP04}
\defcitealias{Pilyugin:2005p308}{PT05}

\defcitealias{Sanchez:2006p331}{San06}
\defcitealias{Moustakas:2006p307}{MK06}
\defcitealias{Ferguson:1998p224}{FGW98}
\defcitealias{Pastoriza:1993p3323}{Pas93}

\begin{document}

\frontmatter

\pdfbookmark[0]{Front matter}{title}
\begin{titlepage}
  
  \thispagestyle{empty}
  \begin{center}
    
    \vspace*{0.07\textheight}
    
    \huge
    \sc
    Multi-dimensional analysis of the\\
    chemical and physical properties\\
    of spiral galaxies

    \vspace*{0.15\textheight}

    \Large
    \rm    
    Thesis submitted for the degree of\\
    Doctor of Philosophy

    \vspace{0.5cm}

    by

    \vspace{1cm}
    
    \Large\sc
    \thesisauthor
    \vspace{0.2cm}
    
    \rm Institute of Astronomy
    \vspace{1cm}

    \includegraphics[width=3.5cm]{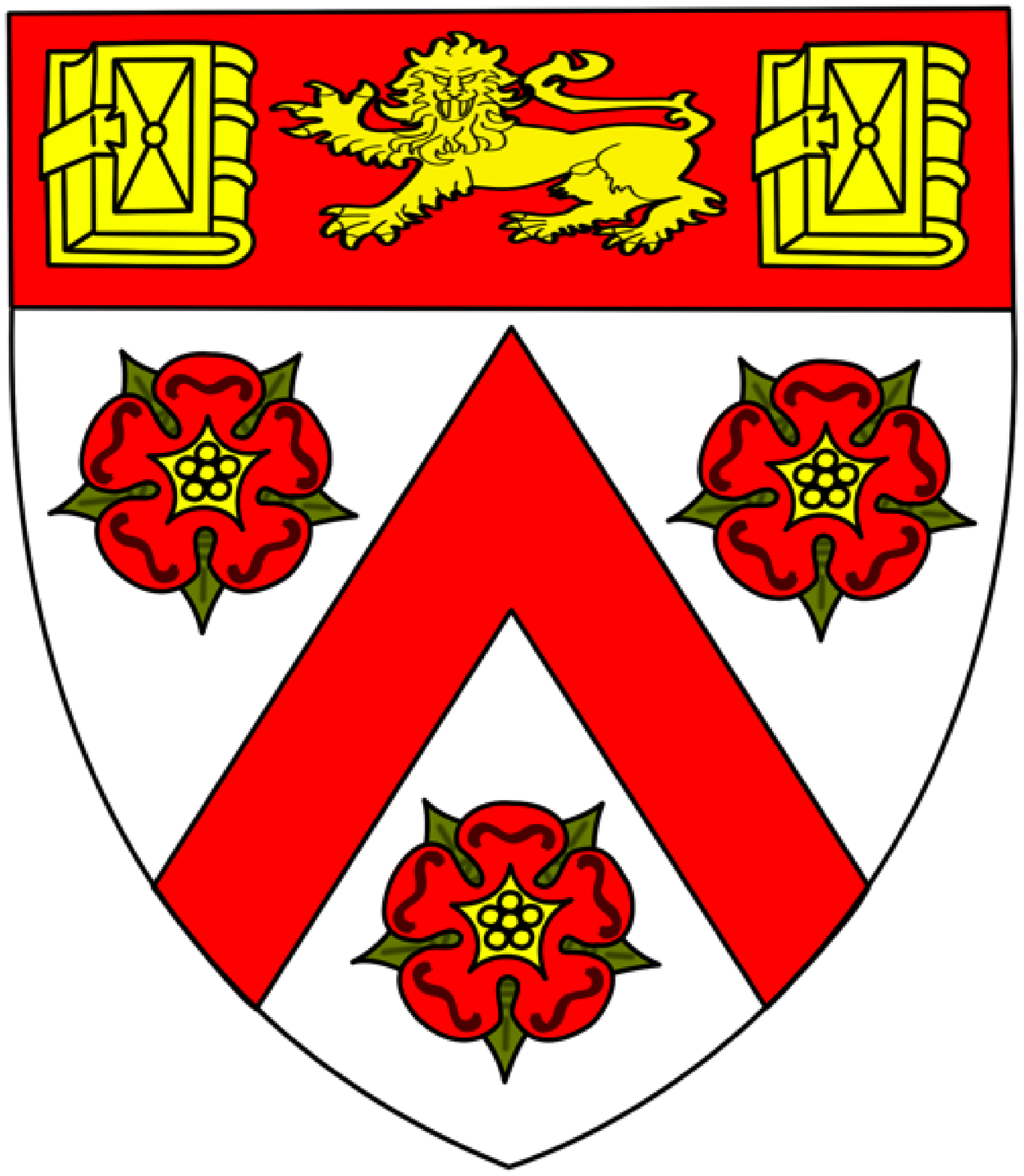}

    {\large\sc Trinity College}
    \vspace{2cm}

   \vspace{0.5\baselineskip}
    University of Cambridge

    \vspace{0.2cm}
    \large
    \thesisdate
    
  \end{center}
  
   \cleardoublepage

\end{titlepage}


\chapter*{Summary}



\vfill

The emergence of a new generation of instrumentation in astrophysics, which
provide spatially-resolved spectra over a large 2-dimensional (2D) field of
view, offers the opportunity to perform emission-line surveys based on
samples of hundreds of spectra in a 2D context,
enabling us to test, confirm, and extend the previous body of results from
small-sample studies based on typical long-slit spectroscopy, while at the
same time opening up a new frontier of studying the 2D structure of
physical and chemical properties of the disks of nearby spiral galaxies.
The project developed in this dissertation represents the first endeavour to
obtain full 2D coverage of the disks of a sample of spiral galaxies in the
nearby universe, by the application of the Integral Field Spectroscopy (IFS)
technique.
The semi-continuous coverage spectra provided by this spectral imaging 
technique allows to study the small and intermediate linear scale variation
in line emission and the gas chemistry in the whole surface of a spiral
galaxy.

The PPAK IFS Nearby Galaxies Survey: PINGS, was a carefully devised
observational project, designed to construct 2D spectroscopic mosaics of 17
nearby galaxies in the optical wavelength range. The sample includes
different galaxy types, including normal, lopsided, interacting and barred
spirals with a good range of galactic properties and star forming
environments, with multi-wavelength public data. The spectroscopic data set
comprises more than 50\,000 individual spectra, covering an observed area of
nearly 100 arcmin$^2$, an observed surface without precedents by an IFS study.
All sources of errors and uncertainties during the 
reduction process of the IFS observations are assessed very carefully. This
methodology contributed not only to improve the standard reduction pipeline
procedure for the particularly used instrument, improvements that can be applied to
any similar integral-field observation and/or data reduction, but to defining
a self-consistent methodology in terms of observation, data reduction and
analysis for the kind of IFS surveys presented in this dissertation, as
well as providing a whole new set of IFS visualization and analysis software
made available for the public domain.


The scientific analysis of this dissertation comprises the study of the
integrated properties of the ionized gas of the whole PINGS sample, and a
detailed 2D study of the physical and chemical abundance distribution derived
from the emission line spectra of four selected galaxies of the sample.
Spatially-resolved maps of the emission line intensities and physical
properties are derived for each the selected galaxies.
Different methodologies are explored in order to study the spatially-resolved
spectroscopic properties of the galaxies. Abundance analysis are performed
based on a variety of diagnostic techniques using reddening corrected spectra.
From this analysis, evidence is found to support
that the measurements of emission lines of a ``classical'' \hh
region are not only aperture, but spatial dependent, and therefore, the
derived physical parameters and metallicity content may
significantly depend on the morphology of the region, on the slit/fibre
position, on the extraction aperture and on the signal-to-noise of the
observed spectrum.
On the other hand, the results presented in this dissertation 
indicate the existence of non-linear {\em multi-modal} abundance gradients in
normal spiral galaxies, consistent with a flattening in the innermost and
outermost parts of the galactic discs, with important implications in terms of
the chemical evolution of galaxies.

The powerful capabilities of wide-field 2D spectroscopic studies are
proven. The chemical composition of the whole surface of a spiral galaxy is
characterised for the first time as a function not only of radius, but of the
intrinsic morphology of the galaxy, allowing a more realistic determination of
their physical properties. The methodology, analysis and results of this
dissertation will hopefully contribute in a significant way to understand the
nature of the physical and chemical properties of the gas phase in spiral
galaxies.


\chapter*{Declaration}

\vspace*{0.5cm}

I hereby declare that this thesis entitled \textit{\thesistitle} is
not substantially the same as any that I have submitted for a degree
or diploma or other qualification at any other University. I further
state that no part of this dissertation has already been or is being
concurrently submitted for any such degree, diploma or other
qualification. This thesis is essentially the result of my own work, and
includes nothing which is the outcome of work done in collaboration except
where specifically indicated. Those parts of this dissertation which are
undergoing review for publication, or included in conference proceedings
are as follows:

\begin{itemize}

\item {\bf Chapter 3}\\
  The work presented in this chapter is essentially my own, it has
  been submitted for publication as:
  Rosales-Ortega, F.~F., Kennicutt, R.~C., Sánchez, S.~F., Díaz, A.~I.,
  Pasquali, A., Johnson, B.~D. and Hao, C.~N., (2009)
  \textit{PINGS: the PPAK IFS Nearby Galaxies Survey},
  submitted to Monthly Notices of the Royal Astronomical Society, and it was
  benefited from collaboration with these authors.

\item {\bf Chapter 4}\\
  The work presented in this chapter is essentially my own, a large
  part of it was submitted for publication in the article mentioned above. The
  work related to NGC\,628 has been partially done within a collaboration, and
  submitted for publication as:
  Sánchez, S.~F., Rosales-Ortega, F.~F., Kennicutt, R.~C.,  Díaz, A.~I.,
  Pasquali, A., Johnson, B.~D. and Hao, C.~N., (2009)
  \textit{PPAK Wide-field Integral Field Spectroscopy of NGC\,628: The largest
  spectroscopic survey on a single galaxy},
  submitted to Monthly Notices of the Royal Astronomical Society. 
  The {\em Gaussian-suppression} and absolute fux calibration techniques were
  developed in collaboration with S.~F. Sánchez. Their implementation and
  initial testing were carried out in parallel by both of us, though the final
  implementation for this dissertation are my own work.

\item {\bf Chapter 5}\\
The work presented in this chapter is essentially my own. Some parts
  has been partially done within a collaboration, and submitted for
  publication in the articles mentioned above.

\item {\bf Chapter 6}\\
  The work presented in this chapter is essentially my own, but benefited from
  discussions and advice from S.~F. Sánchez and A.~I. Díaz.

\item Some figures of chapter 3 and chapter 6 have been included in 
  \textit{The Promise of Multiwavelength and IFU observations},
  Kennicutt, R.~C., Hao, C.~N., Johnson, B.~D., Rosales-Ortega, F.~F., Díaz,
  A.~I., Pasquali, A. and Sánchez, S.~F., (2009) Proceedings of the IAU Symposium
  262, G.~Bruzual, S.~Charlot, eds.

\end{itemize}

This thesis is less than 60\,000 words in length.

\vspace{0.5cm}
\parbox{10cm}{
  \thesisauthor \\
  \footnotesize{Cambridge, \thesisdate}
}



\vspace*{1cm}

\chapter*{Thesis content}

\vspace*{1cm}

\begin{center}

\Large
The whole thesis is not included in astro-ph due to file size limitations.\\
\vspace*{1cm}

The full contents can be found at:\\

\vspace*{0.5cm}

\url{http://www.dspace.cam.ac.uk/handle/1810/224843}

\end{center}


\mainmatter

\setcounter{equation}{0}



\setcounter{chapter}{0}

\chapter[Introduction]{Introduction}
\label{chap:1}

\dropping{4}{T}he existence and distribution of the chemical elements and
their isotopes in the universe is a consequence of very complex processes that
have taken place in the past since the Big Bang and subsequently in stars and
in the interstellar medium (ISM) of the present day galaxies, where they
are still ongoing. These processes have been studied theoretically,
experimentally and observationally. Different theories of cosmology, stellar
evolution and interstellar processes have been considered, laboratory
investigations of nuclear and particle physics, studies of elemental and
isotopic abundances in the Earth and meteorites have also been involved, as
well as astronomical observations of the physical nature and chemical
composition of stars, galaxies and the interstellar medium.

From the observational point of view, the study of chemical abundances in galaxies, like
many other areas of astrophysics, has undergone a remarkable acceleration in
the flow of data over the last few years. 
We have witnessed wholesale abundances determinations in tens of
thousands of galaxies from large scale surveys such as the Two Degree Field
Galaxy Redshift Survey \citep[2dFGRS,][]{Colless:2001p2675} and the Sloan
Digital Sky Survey \citep[SDSS,][]{York:2000p2677},
measurements of abundances in individual stars of Local Group galaxies beyond
the immediate vicinity of the Milky Way, and the determination of the chemical
composition of some of the first stars to form in the Galactic halo. Chemical
abundances studies are also increasingly being extended to high redshift,
charting the progress of stellar nucleosynthesis over most of the age of the
universe. The primary motivation common to all of these observational efforts is
to use the chemical information as one of the means at our disposal to link
the properties of high redshift galaxies with those we see around us today, and
thereby understand the physical processes at play in the formation and
evolution of galaxies.

The galactic chemical evolution is dictated by a complex
array of parameters, including  the local initial composition, star formation
history (SFH), gas infall and outflows, radial transport and mixing of gas
within disks, stellar yields, and the initial mass function (IMF). Although is
difficult to disentangle the effects of the various contributors, measurements
of current elemental abundances constrain the possible evolutionary histories
of the existing stars and galaxies.
Important constraints on theories of galactic chemical evolution and on the star
formation histories of galaxies can be derived from the accurate determination
of chemical abundances either in individual star-forming regions distributed
across galaxies or through the comparison of abundances between galaxies. 
Nebular emission lines from individual \hh regions have been,
historically, the main tool at our disposal for the direct measurement of
the gas-phase abundance at discrete spatial positions in low redshift
galaxies. 

However, in order to obtain a deeper insight of the mechanisms that rule the
chemical evolution of galaxies, we require a significantly the number of \hh
regions sampled in any given galaxy. 
In this dissertation, I present a new observational technique
conceived to tackle the problem of the 2-dimensional coverage of the whole
surface of a galaxy. The advent of new spectroscopic techniques provides powerful
tools for studying the small and intermediate scale-size variation in line
emission and stellar continuum in nearby well-resolved galaxies. In this work, I
address the problems and challenges that imply the determination of the
chemical composition in galaxies in a 2D context and the subsequent derivation
of their physical properties. 

I will begin by presenting in this chapter a literature review on the
determination of chemical abundances in galaxies. As an introduction to this
topic, the physics of gaseous nebulae is discussed in \S\,1.1, together with
a discussion of extra-galactic \hh regions in \S\,1.2.
The different methods of abundance determinations are presented in \S\,1.3. 
Physical properties derived from the determination of chemical abundances 
are discussed in \S\,1.4. These latter sections are
partially based on the paper reviews and books about the physics and chemistry of
the interstellar medium and \hh extragalactic regions by
\citet{Dinerstein:1990p1723}, \citet{PerezMontero:2005p306},
\citet{Tielens:2005p1750}, and \citet{Osterbrock:2006p2331}.
This discussion leads to the presentation of new techniques and
methods for the determination of chemical abundances in nearby galaxies as
described in \S\,1.5


\section{The Physics of Gaseous Nebulae}

Gaseous Nebulae are observed as bright extended objects in the sky, some are
easily observed on direct images but many others are intrinsically less luminous
or are affected by interstellar extinction on ordinary images, but can
be resolved on long exposures with special filters and techniques so that the
background and foreground stellar and sky radiations are suppressed. The surface
brightness of a nebula is independent of its distance, but more distant nebulae
have (on average) smaller angular size and greater interstellar extinction.

Gaseous nebula have emission-line spectra. Nebulae emit electromagnetic
radiation over a broad spectral range, although only
a few wavelengths pass easily through the Earth's atmosphere. Visible light and
some infrared and radio radiation can be studied from the ground, but most other
wavelengths can only be covered from high-latitude aircrafts or space
telescopes.
The source of energy that enables normal emission nebulae to radiate is
ultraviolet radiation from stars within or near the nebula. There should be one
or more stars with effective surface temperature $T_{\star} \ge$ 3 x 10$^4$ K, 
the ultraviolet photons of these stars transfer energy to the nebula by
photoionization. In all nebulae, hydrogen (H) is by far the most abundant element, and
photoionization of H is thus the main energy-input mechanism. Photons with
energy greater than 13.6 eV (the ionization potential of H), are absorbed in the
process, and the excess energy of each absorbed photon over the ionization
potential appears as kinetic energy of a newly liberated photoelectron. Collisions
between electrons and between electrons and ions, distribute this energy and
maintain a Maxwellian velocity distribution with temperature $T$ in the range
5,000 $<$ $T$ $<$ 20,000 K in typical nebulae. Collisions between thermal
electrons and ions excite the low-lying energy levels of the ions. Downward
radiation transitions from these excited levels have very small transition
probabilities, but at the low densities ($n_e \leq$ 10$^4$ cm$^{-3}$) of typical
nebulae, collisional de-excitation is even less probable, so almost every
excitation leads to emission of a photon, and the nebula thus emits a
forbidden-line spectrum that is quite difficult to reproduce under terrestrial
laboratory conditions.

Thermal electrons are recaptured by the ions, and the degree of ionization at
each point in the nebula is fixed by the equilibrium between photoionization and
recapture. In the recombination process, recaptures occur to excited levels, and
the excited atoms thus formed then decay to lower and lower levels by radiative
transitions, eventually ending in the ground level. In this process, line
photons are emitted and this is the origin of the \hi Balmer and Paschen line
spectra observed in all gaseous nebulae. The recombination of H$^+$ gives rise
to excited atoms of H$^0$ and thus leads to the emission of the \hi
spectrum. Likewise, He$^+$ recombines and emits the \hei spectrum, and in the
most highly ionized regions, He$^{++}$ recombines and emits the \heii
spectrum. Recombination lines of trace elements are also emitted; however, the
main excitation process responsible for the observed strengths of such lines
with the same spin or multiplicity as the ground term is resonance fluorescence
by photons, which is much less effective for H and He lines because the
resonance lines of these more abundant elements have greater optical
depth. Nevertheless, line emission of these rare elements plays a significant
role in the physics of the nebula, and permits the determination of the
chemical composition inside the nebula.

The spectra of gaseous nebulae are dominated by collisionally excited
forbidden lines of ions of
common elements, such as \oiii \lam\lam4959,\,5007 (the famous
green nebular lines); \nii \lam\lam6548,\,6584 and \sii
\lam\lam9069,\,9523 in the red; and \oii
\lam\lam3727, 3729 in the ultraviolet (which normally appears as a
blended \lam3727 line on low-dispersion spectrograms). In addition, the
permitted lines of hydrogen, H$\alpha$ \lam6563 in the red, H$\beta$
\lam4861 in the blue, H$\gamma$ \lam4340 in the violet and so on, are
characteristic features of every nebular spectrum, as is \hei \lam5876,
which is considerably weaker, while \heii \lam4686 occurs only in
higher-ionization nebulae. Long-exposure spectrophotometric observations
extending to faint intensities show progressively weaker forbidden lines, as
well as faint permitted lines of common elements such as C\,{\footnotesize
  II}, C\,{\footnotesize III}, C\,{\footnotesize IV} and
so on. Nebular emission-line spectra extend into other spectral ranges,
in the infrared for example, the [Ne\,{\footnotesize II}] \lam12.8 $\mu$m
and \oiii \lam88.4 $\mu$m are among the strongest lines measured, into the
ultraviolet Mg\,{\footnotesize II} \lam\lam2796,\,2803, C\,{\footnotesize III}]
\lam\lam 1907,\,1909, C\,{\footnotesize IV}
\lam\lam1548,\,1551 and Ly$\alpha$ \lam1216 are also
observed. Is often necessary to obtain spectra outside the traditional
visible/near spectral bands to get an accurate picture of the system in
question.

Gaseous nebulae have weak continuous spectra, consisting of atomic and
reflection components. The atomic continuum is emitted chiefly by free-bound
transitions, mainly in the Paschen continuum of \hi at \lam $>$ 3646 \AA,
and the Balmer continuum at 912\,\AA\, $<$ \lam $<$ 3646\,\AA. In addition to
the bright-line and continuous spectra emitted by atomic processes,
many nebulae have reflection continua arising from starlight scattered by
dust. The amount of dust varies from nebula to nebula, and the strength of this
continuum fluctuates correspondingly. In the infrared for example, the nebular continuum is
largely thermal radiation emitted by dust particles heated to a temperature of
order 100 K by radiation derived originally from the central star.

Gaseous nebula may be classified into two main types: \hh regions and planetary
nebulae (PNe). Though the physical processes in both types are quite similar, the two
groups differ greatly in origin, mass, evolution and age. The objects of study
for the chemical composition of galaxies are extragalactic \hh regions. These
diffuse nebulae are regions of interstellar gas in which the exciting stars are
O- or early B-type stars, i.e. young stars which use up their nuclear energy
quickly. These hot, luminous stars undoubtedly formed fairly recent from
interstellar matter that would be otherwise be part of the same nebula. 
The effective temperature of the stars are in the range 3 x 10$^4$ $<$ $T_{\star}$
$<$ 5 x 10$^4$ K; throughout the nebula, H is ionized, He is singly ionized and
other elements are mostly singly or doubly ionized. Typical densities in the
ionized part of the nebula are of the order 10 or 10$^2$ cm$^{-3}$, ranging to
high as $10^4$ cm$^{-3}$. In many nebulae, dense neutral condensations are
scattered through the ionized volume. Internal motions occur in the gas with
velocities of order 10 km\,s$^{-1}$, approximately the isothermal sound
speed. Bright rims, knots, condensations, and so on, are apparent to the limit
of resolution. The hot, ionized gas tends to expand into the cooler surrounding
neutral gas, thus decreasing the density within the nebula and increasing the
ionized volume. The outer edge of the nebula is surrounded by ionization fronts
running out into the neutral gas. This original two-phase model of the
interstellar medium (the \hh/\hi region dichotomy) was introduced by
\citet{Stromgren:1939p1753}.
He showed that photoionized gas near hot stars is segregated into physically
distinct volumes, separated from their neutral environment by sharp boundaries.


\subsection{Extragalactic \hh regions}

The spectra of \hh regions are strong in \hi recombination lines, \nii, \oii
and \oiii collisionally excited lines, but the strengths of N and O
may differ greatly, being stronger in the nebulae with higher central-star
temperatures. The brightest \hh regions can easily be seen on almost any
large-scale image of nearby galaxies, and those taken in a narrow wavelength
band in the red (including H$\alpha$ and \nii lines) are specially effective
in showing faint and often heavily obscured extragalactic \hh regions. The \hh regions are
strongly concentrated to the spiral arms and indeed are the best objects for
tracing the chemical composition, structure and dynamics of the spiral arms in
distant galaxies. They trace recent star formation and, through the analysis
of their chemical composition, previous star formation activity. Typical
masses of observed \hh regions are of the order 10$^2$ to 10$^4$ M$_{\odot}$,
with the lower limit depending largely on the sensitivity of the observational
method used.

\hh regions are the only form of interstellar material which emits strongly in the 
optical spectral region; therefore, there is a much longer and richer history of
observations and theory for them than for the other thermal phases of
interstellar matter. 
Optical observations of \hh regions provide fairly complete information about 
their elemental composition. From their spectra, abundances relative to hydrogen 
can be estimated for nearly all of the most common elements, particularly He, N, 
O, Ne, Ar, and S (note that oxygen alone constitutes nearly 50\% by mass of the 
elements heavier than helium.) Furthermore, ionized nebulae are remarkably 
efficient machines for converting ultraviolet continuum energy from OB stars, 
originally diluted over wide bandpasses, into a few narrow, intense,
optically-thin emission lines. The intrinsic emissivities of these lines are
easy to calculate in principle, although they are sensitive to the local
thermodynamic state of the gas (electron density $n_e$ and temperature $T_e$). 
On the other hand, the thermal parameters can also be determined from the
spectra, using diagnostic line-intensity ratios. In this way, \hh regions can
be used to measure element abundances in the (present-day) gas of distant
galaxies.

The sample of extragalactic \hh regions studied so far has metal abundances
ranging from about 0.02 to several times solar. This is a useful complement to
studies of our own Galaxy, which contains no severely metal-deficient \hh
regions (except for a handful of planetary nebulae formed by stars of the halo
population). In contrast, for many \hh regions in the outskirts of late-type
spirals and in some  dwarf irregular galaxies, the process of metal enrichment
by stellar nucleosynthesis is still in its early stages, providing a hint on the early
chemical evolution of galaxies. These low-metallicity \hh regions are also
presumed to have experienced only a small degree of  alteration in their
helium abundances due to stellar activity. Therefore, their present He/H
ratios should be nearly the same as the primordial value, providing valuable
tests for cosmological theories. The various {\em categories} of extragalactic
\hh regions are essentially lists of their environments. These include: 

\begin{enumerate}
\item Disk \hh regions in spiral and irregular galaxies.
\item Gassy dwarf irregular galaxies with spectra which are heavily dominated
  by \hh regions.
\item Nuclear and near-nuclear regions sometimes called ``starburst'' or
  ``hotspot'' \hh regions \citep[e.g.][]{Kennicutt:1989p1756}.
\end{enumerate}

The first two categories have the best abundance data available in the
literature. Regions in the third group tend
to have relatively strong stellar continua and to be fairly metal-rich, which
make it to difficult to obtain accurate measurements of the emission lines
from which abundances are determined. On the other hand, members of the first
two categories are universally regarded as members of the same family. \hh
regions in nearby galaxies have been well-catalogued; atlases are available
for the Large and Small Magellanic Clouds (LMC, SMC), and a large number of
other galaxies \citep[e.g.][]{Hodge:1967p1761,Hodge:1977p1770,Hodge:1983p1782}. The
star-forming dwarf irregulars are usually found by spectroscopic surveys for
emission-line galaxies (e.g. \citealt{Kinman:1984p1785} and more recently the
SDSS data releases).

The statistical properties of the \hh region populations in spiral and
irregular galaxies were addressed by \citet{Kennicutt:1988p1791} and
\citet{Kennicutt:1989p1793}. They find that late-type galaxies have both 
intrinsically higher-luminosity \hh regions, and larger total
numbers of \hh regions after normalization by galaxy size, than do early-type
spirals. 
Within a galaxy, the differential luminosity function of the \hh regions is
roughly a power-law, $N\, \propto L^{-2\,\pm\,0.5}$, although some
low-luminosity irregulars have an exceptional supergiant complex, and Sa-Sb
galaxies are deficient in luminous regions. While the positive correlation
between the luminosity of the brightest \hh region and that of the parent
galaxy can be understood as chiefly a sample-size effect, the dependence on
morphological type is a real and separate factor. Typical large galaxies
contain hundreds of optically detectable \hh regions. It is important to
note that of all the regions detected and cataloged in H$\alpha$ or H$\beta$,
it is usually the nearest and the most luminous ``giant'' \hh regions for
which abundances are derived.

Some of the best-studied regions are the 30 Doradus complex in the LMC, NGC\,604
in M\,33, and NGC\,5461 and 5471 in M\,101. Selection effects play an
important role, necessarily poorer spatial resolution contributes to a
tendency to identify larger regions in more distant galaxies. This effect is
illustrated by \citet{Israel:1975p1797}, who compare large-beam radio
measurements with optical images of the same \hh regions in M\,101; at better
resolution these regions break up into groups or chains of smaller
clumps. Likewise, \hh regions in dwarf irregulars are also found to have
complex structure when closely examined
\citep[e.g.][]{Hodge:1989p1804,Davidson:1989p1815}. In more distant galaxies,
we will always be looking at more heterogeneous volumes; for
example, a typical aperture size (4'') for spectrophotometric studies
corresponds to 1 pc at 50 kpc (the LMC) and 2 kpc at 100 Mpc.


The morphology of many giant extragalactic \hh regions can be characterized 
to first order as a ``core-halo'' structure, on the basis of both optical and 
radio-continuum data. The cores are composed of dense material, often in 
several distinct clumps, close to the ionizing stars. The diffuse, lower-density 
envelopes are presumably ionized by photons escaping from the inner regions 
and represent the radiation-bounded edges of the Str{\"o}mgren volume. Most giant 
extragalactic \hh regions are believed to be essentially radiation-bounded
\citep[e.g.][]{McCall:1985p1243}. In addition, the denser regions themselves
are inhomogeneous, as seen in the detailed studies of NGC\,5471 by
\citet{Skillman:1985p1843}, and of NGC\,604 by \citet{Diaz:1987p1845}. That
there are also inhomogeneities on smaller spatial scales is shown by the
discrepancy between (rms) $n_e$ values derived from recombination emission and
local values determined from density-sensitive line ratios. The dense clumps
are embedded in a much lower-density medium, with typical clump volume filling
factors of 0.01\,--\,0.1
\citep[e.g.][]{Kennicutt:1984p1848,McCall:1985p1243}. The interclump material
is often treated as a vacuum in nebular models, because it does not contribute
significantly to the optical emission lines.

Giant extragalactic \hh regions display supersonic velocities, which appear to
correlate with H$\beta$ luminosity. 
\citet{Terlevich:1981p1855} interpret the line-widths as virial
and therefore usable for determining the local gravitational field; they also
find a secondary dependence on metallicity. An alternative interpretation of
the origin of the line-widths is that they are a result of stellar winds from
the exciting stars, and possibly also from embedded supernova remnants
\citep[e.g][]{Dopita:1981p1882,Skillman:1985p1843}. For nearby regions, it is
possible to actually identify the stars which may be responsible for driving
the high-velocity gas.

As mentioned above, luminous extragalactic \hh regions are ionized by OB associations. 
For nearby regions, the members of the stellar cluster can be distinguished 
individually and HR diagrams can be constructed. The nebular ionization
structure and emitted spectrum will evolve as the cluster ages and the UV
radiation field diminishes and softens. 
Wolf-Rayet stars are often present in extragalactic \hh
regions, the frequency of Wolf-Rayet stars is higher for higher-metallicity
regions as proved by \citet{Maeder:1980p1901}. Wolf-Rayet stars are important in this 
context because they furnish metal-rich outflows which are capable of altering 
the chemical composition of their gaseous environment. Giant \hh regions are
also known hosts of Type II supernovae. Winds and supernovae from the massive
stars can contaminate (or enrich) the local gas in \hh regions in He, C, O,
and other species. Evidence for such local enrichments has been sought and
perhaps seen in some regions \citep{Skillman:1985p1843, Pagel:1986p1913}.



\section{Determination of chemical abundances in \hh regions}

\hh regions are ideal places to determine the abundance of the elements that
are responsible for recombination and fine-structure lines. The list of these
elements is generally limited at the present time, although lines of many more
elements are observable in several planetary nebulae and high spectral
resolution and multi-wavelength studies of nearby \hh regions
\citep[e.g.][]{GarciaRojas:2006p2832}.
The determination of element abundances in \hh regions are given relative to the
hydrogen content, which is observed by its recombination lines. Only a few abundant
elements give observable recombination lines with similar physics: helium,
carbon, nitrogen and oxygen, whose lines are very weak to detect and may suffer
problems with fluorescence excitation. The abundances derived from the
fine-structure lines in the visible are sensitive to both temperature and
density, and the interpretation of the line intensities is a delicate
problem. In some cases, the temperature of the emitting zone can be obtained
and the abundance determination is safer. However, even in this case
temperature fluctuations can yield systematic errors in the abundances.

The abundances derived from the mid-and far-infrared fine structure lines are
not sensitive to electron temperature and are little affected by
extinction. These are considerable advantages with respect to the optical
lines. However, there are important discrepancies between the abundances
derived from infrared and from optical lines. These differences may originate
in temperature fluctuations or in errors in atomic parameters, but one has to
consider that the critical density for infrared lines is generally much
smaller than for visible lines, so that the abundances derived from the
infrared lines are underestimated if the density is high (this effect can be
important for planetary nebulae).

\begin{figure}[!t]
  \begin{center}
    \includegraphics[width=10cm]{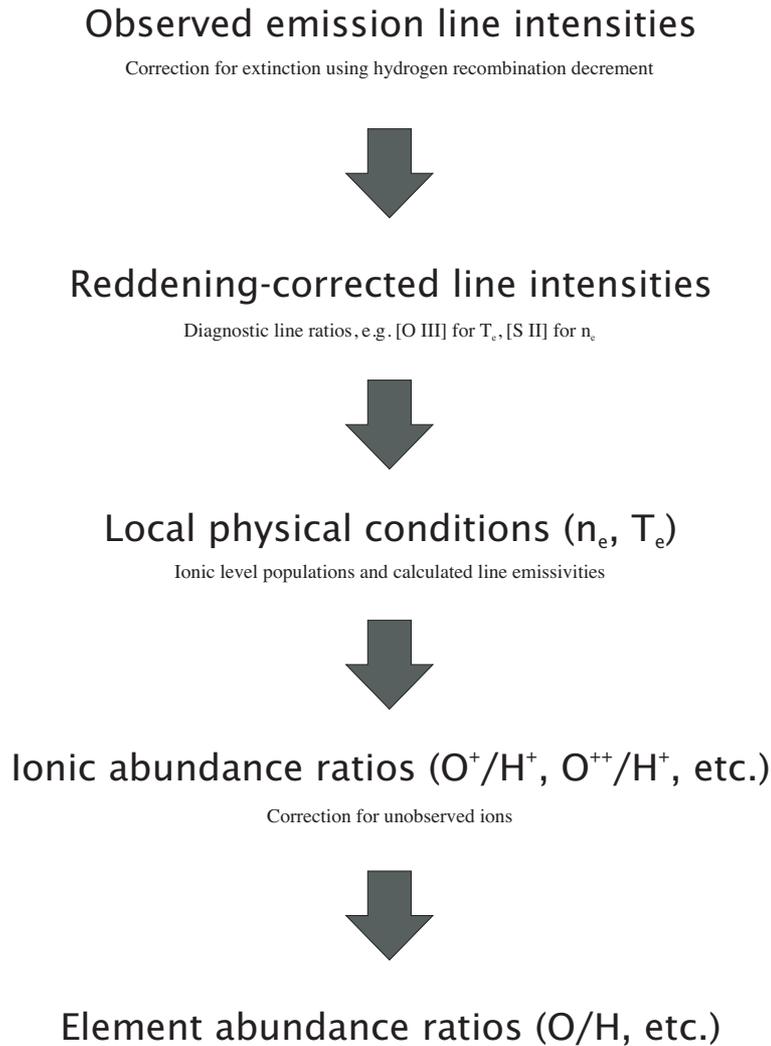}
    \caption[The {\em direct} method flow-chart]{The {\em direct method} of chemical abundance
      determinations.}
    \label{fig:direct}
  \end{center}
\end{figure}

All elements (with the obvious exception of H) exist in several ionization
states in \hh regions. However, only the abundances of those ions that emit
observable lines can be determined. If such ions are minor species, they yield
no useful information because the physical parameters of \hh regions are most
often too uncertain to allow an accurate solution of the ionization
equilibrium. This is for example the case for O\,{\footnotesize I},
C\,{\footnotesize II}, S\,{\footnotesize II} or Si\,{\footnotesize II}. If the
observed ion is a major species the situation is more favorable since we can
calculate, more or less accurately, the abundances of the unobserved ions of
the same element. However, uncertainties remain if a high precision is
required, as is the case of helium in a cosmological context. The most
favorable case is that of oxygen, whose major ionization states,
O\,{\footnotesize II} and O\,{\footnotesize III}, are observable optically and
for which the electron temperature $T_e$ can be determined. For this
reason, oxygen is, after helium, the element whose abundance is best
determined, at least if the temperature is large enough (or the metallicity is
too low) for the
temperature-sensitive lines (e.g. \oiii \lam4363) to be measured. If this is not the
case, we may construct tailored models of the nebular ionization and thermal
structure of a \hh region to estimate the electron temperatures and 
ionization correction factors for individual ions.

However, given the difficulty of detecting the $T_e$-sensitive
line and the assumptions made in nebular modeling, a very popular approach is to
obtain the abundance of extragalactic \hh regions using empirical relations
between the oxygen abundance and the intensity of the \oii
$\lambda\lambda$3726,\,3729 and \oiii $\lambda\lambda$4959,\,5007 lines
relative to H$\beta$
\citep{Pagel:1997p2833} or by using the \oii $\lambda\lambda$7320,\,7330 as 
described by \citet{Aller:1984p1918} and implemented by
\citet{Kniazev:2004p340} in SDSS \hh galaxies. This method however, is the
less accurate and much discussion about the reliability of the different
empirical calibrations is still ongoing in the literature \citep[e.g. see][for a
thorough discussion]{Kewley:2008p1394}. A full discussion regarding this topic
is beyond the scope of this chapter, however, in \autoref{chap:5},  I include
a small review on the different empirical techniques of abundance
determinations (considering their particular advantages and pitfalls), and
their implementations in the context of the work carried out in this
dissertation. A more complete explanation of the
determination of nebular abundances from emission lines can be found in
references on the physics of gaseous nebulae such as \citet{Aller:1984p1918},
and \citet{Osterbrock:2006p2331}.

\section{Abundance gradients in galactic disks}

It has been noticed that certain \hh region emission-line ratios, such as
[O\,{\footnotesize III}]/H$\beta$, vary across the disks of nearby spiral galaxies. The
interpretation of this variation in terms of a metallicity trend was introduced
by \citet{Searle:1971p1962}, in a paper that laid the groundwork for the entire field of
abundance gradients. It was soon followed up by further observational studies
and a more rigorous analysis involving the construction of realistic nebular
models \citep{Shields:1974p1960}. From the start it was recognized that
there was a need for a ``second parameter'' in addition to the O/H ratio, to
explain an observed systematic increase in O$^{++}$/O$^+$ with decreasing
O/H. \citet{Shields:1976p2006} suggested that this secondary effect results
from a tendency for the effective temperatures of the ionizing stars to be
hotter for lower O/H, and interpreted it as a metallicity-dependent truncation
of the top end of the initial mass function (i.e. that the formation of very
massive stars is inhibited by higher metallicity). Some form of the idea of a
$Z$-dependent IMF is still a popular interpretation of the ``excitation'' trend 
\citep[e.g.][]{Vilchez:1988p2011}, but it is also the case that a similar
effect can arise from systematic variations in the nebular geometry and/or
filling factor \citep{Mathis:1985p1963,Dopita:1986p1586}.

An extensive body of literature has been amassed on the subject of abundance
gradients in galaxies. Not surprisingly, many works have focused on large,
nearby galaxies with many observable \hh regions, such as M\,33
\citep{Vilchez:1988p2018, Rosolowsky:2008p1636} and M\,101 
\citep{Evans:1986p1969, TorresPeimbert:1989p2005,Kennicutt:1996p1603}. The gradients are usually
expressed as a logarithmic fit to some 5\,--\,20 regions per galaxy, and have a
magnitude of about $\delta$log(O/H)/R\,=\,-0.08 ($\pm$ 0.03) dex/kpc. This is
similar to the values derived for the solar-neighborhood metallicity gradient
in the Milky Way galaxy. 
The trend of these gradients in the inner parts of galactic disks are
difficult to study, both because the \hh region samples are often small, and
more fundamentally because these are generally the most metal-rich regions,
for which \oiii \lam4363 is unobservable and therefore the derived
abundances are heavily model-dependent.

The steepest abundance gradients were initially seen in late-type spiral
galaxies (types Sb-Scd). Irregulars and barred spirals tend to have weak or zero
radial gradients. Early type spirals are harder to study because their \hh
regions are intrinsically fainter, but studies of M\,81 (Sab) show it to have 
an O/H gradient similar to those of M\,33 and M\,101 \citep{Garnett:1987p226}. 
There is at present no convincing evidence that the O/H gradient depends
on morphological type among spiral galaxies. However, there is evidence for a
good correlation between mean O/H abundance and the overall galaxy mass or
luminosity. This trend resembles the correlation of stellar metallicity with
galaxy mass, and probably has its roots in the fundamental processes of galaxy
formation and evolution.

\begin{figure}[!t]
  \begin{center}
    \includegraphics[width=\textwidth]{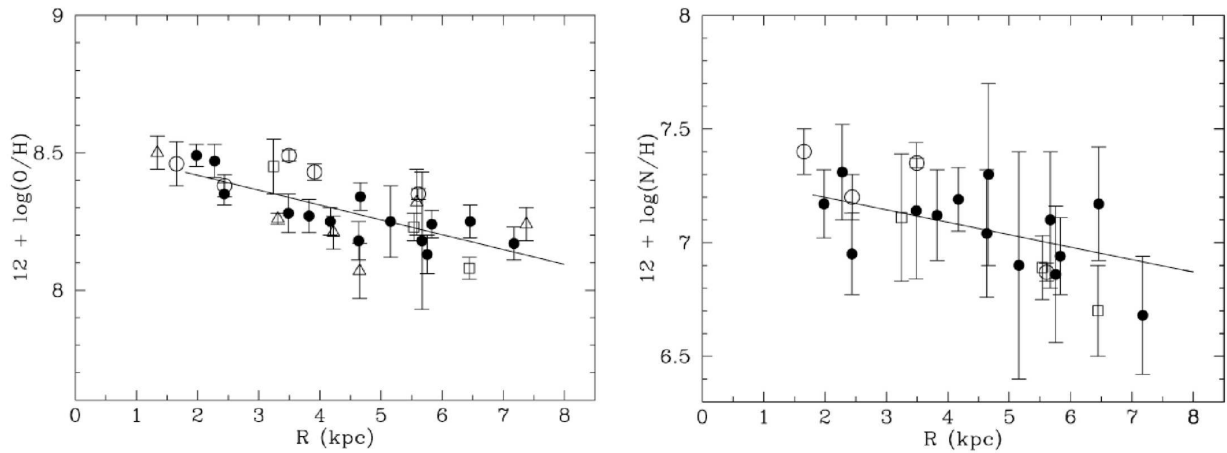}
    \caption[Radial abundance gradients of M\,33.]{The O/H and N/H abundance
      vs. galactocentric distance in M\,33, examples of the radial oxygen and
      nitrogen abundance gradients. Plots taken from \citet{Magrini:2007p200}.}
    \label{fig:gradient}
  \end{center}
\end{figure}

Along with the trend in [O\,{\footnotesize III}]/H$\beta$, a similar radial trend was noted
for the ratio [N\,{\footnotesize II}]/H$\alpha$, which decreases with
increasing distance from the centers of spiral galaxies. Although part of this
trend is due to the generally lower degree of ionization in the outer \hh
regions, there also must be a real variation in abundance. 
Unlike oxygen, for nitrogen one usually can measure the singly-ionized state
only; unfortunately, N$^{++}$ has no strong optical lines. As mentioned
before, the nitrogen abundance is basically derived from [N\,{\footnotesize
  II}]/[O\,{\footnotesize II}]. The
relative behavior of O and N is often displayed by plotting N/O vs. O/H. 
Some studies find that N/O varies almost as steeply as O/H, which has special
significance in the context of chemical evolution models, but others claim that N/O varies
only slightly or is constant across the disks of galaxies such as M\,101, M\,33,
M\,81, and M\,83. There also appear to be variations in N/O
at a given O/H from galaxy to galaxy (same references as above). Some of
these variations may be an artifact of the analysis, especially since N$^+$
contains only a small fraction of the nitrogen for the lowest-abundance, most
highly ionized regions. For such regions, the ionization correction factors 
are very large, and the uncertainties in the ionization
structure translate into large uncertainties in the elemental abundance of
nitrogen. Nevertheless, there is accumulating evidence that nitrogen has a more
complicated behavior than does oxygen, with N/O being roughly constant at low
values of O/H and increasing at higher O/H
\citep[e.g.][]{Pagel:1985p2022,TorresPeimbert:1989p2005}. Measurements of N/O
in metal-poor dwarf irregular galaxies are an important ingredient in this argument.

Scatter in gradient determinations has been seen in various studies
(e.g., in the Milky Way \citealt{Afflerbach:1997p2977} or in M\,33
\citealt{Rosolowsky:2008p1636}), even after accounting for uncertainties in
the stellar absorption and reddening corrections, an intrinsic scatter of $\sim$
0.1 dex has been measured in these very well-studied galaxies which is
unexplained by the measurement uncertainties. Regardless of its source,
gradient determinations made in the face of significant scatter coupled with a
limited number of observations may produce widely varying results. This
historical evolution of the gradient determination ranging over nearly an
order of magnitude, should serve as a cautionary example.
Only large numbers of measurements can overcome the uncertainties
engendered by the intrinsic variance, as some observations suggest
that the uncertainties in the gradients are systematically underreported.


\subsection{The Galactic abundance gradient}

Because of interstellar extinction, one can use the same techniques as for
extragalactic \hh regions only for the part of our Galaxy outside a
galactocentric distance of about 7 kpc. Studies such as those by \citet{Hawley:1978p2030}
 found gradients similar to those in other spirals, $\delta$log(O/H)/$\delta$R
= -0.04 to -0.06 dex/kpc and $\delta$log(N/H)/$\delta$R = -0.10 dex/kpc. Determination of
abundances in the inner galaxy requires the use of other techniques, such as
measuring electron temperatures from radio recombination lines. The values of $T_e$
are found to increase systematically with increasing radius, presumably because of a
decreasing abundance of oxygen, the primary coolant. The inferred gradient in
O/H  is $\delta$log(O/H)/$\delta$R = -0.07 dex/kpc after the classic paper of
\citet{Shaver:1983p2040}.

The results from optical studies for the other measurable elements are similar
to those for other galaxies: N/H varies more steeply than O/H; 
S/O, Ne/O, and Ar/O do not vary in the outer part of the Galactic disk. Again,
the optical studies are restricted to the unobscured portion of the Milky Way
galaxy, and therefore do not sample the inner disk where the inferred O/H
values are high. A more recent development, made possible by improvements in
infrared detectors and the availability of space observatories. 
The exploration of the infrared spectral region as a tool for studying
the galactic abundance gradient. The
mid-infrared spectral region (5-30 $\mu$m) contains emission lines of the major ions
of Ar, S, and Ne: [Ar\,{\footnotesize II}] 7.0 and [Ar\,{\footnotesize III}]
9.0 $\mu$m; [S\,{\footnotesize III}] 18 and [S\,{\footnotesize IV}] 10.5
$\mu$m; and [Ne\,II] 12.8 $\mu$m. These lines have been measured in a number of \hh
regions in the inner Galaxy, and evidence for abundances elevated by factors of
two or three have been found for the Galactic Center and
for \hh regions in the 5 kpc ``ring'' region \citep{Pipher:1984p2082}.

However, even these mid-infrared lines suffer somewhat from extinction. In
particular, the [Ar\,{\footnotesize III}] and [S\,{\footnotesize IV}] lines
fall in the middle of the strong 10 $\mu$m
silicate absorption feature, where the optical depth is comparable to that in
the near-infrared. Another approach to studying abundances in the inner galaxy
is to make use of the fine-structure lines of \oiii 52, 88 $\mu$m and [N\,III] 57
$\mu$m. By a happy coincidence, these lines from the abundant and (presumably)
usually co-extensive O$^{++}$ and N$^{++}$ ions fall close together in wavelength and have
fairly similar dependences on the electron density. The line emissivities are
also essentially independent of the electron temperature. Measurements of these
three lines therefore yield a relatively accurate value for the N/O ratio. 
A survey of about a dozen galactic H
II regions in these lines yielded strong evidence that N/O in the Galactic
Center and 5 kpc ``ring'' is elevated by a factor of 2 or 3 as compared to the
solar neighborhood. There remain some unsettled questions
regarding N/O determinations from the far-infrared lines, including possible
ionization structure effects in \hh regions ionized by very cool stars and a
systematic discrepancy between values derived from the infrared lines and
those derived optically from [N\,{\footnotesize II}]/[O\,{\footnotesize II}]
\citep{Rubin:1988p2103}. 

More recent observations of IR fine-structure lines of the [S\,{\footnotesize
  III}] 19 $\mu$m, [O\,{\footnotesize III}] 52 and 88 $\mu$m, and
[N\,{\footnotesize III}] 57 $\mu$m in compact \hh
Galactic regions have found abundance gradients of the form [S/H] = (-4.45
$\pm$ 0.04) - (0.063 $\pm$ 0.006) (kpc), [N/H] = (-3.58 $\pm$ 0.04) - (0.072
$\pm$ 0.006) (kpc), and [O/H] = (-2.85 +/- 0.06) - (0.064 +/- 0.009) (kpc)
\citep{Afflerbach:1997p2977}. These abundances are consistent with
production of sulphur, nitrogen, and oxygen by primary
nucleosynthesis. Comparison with abundances in other galaxies implies a
Hubble type between Sab and Sb for our Galaxy and an unbarred or mixed
galactic structure \citep{VilaCostas:1992p322}.

\section{Comparison with Chemical Evolution Models}

The recognition of significant variations in the gas composition within and
among galaxies, along with parallel results on the stellar populations, inspired
the development of chemical evolution models which attempt to explain these
patterns. The so-called ``simple model'' postulates a closed system of gas and
stars, which self-enriches in metals as generations of stars age, die, and seed
the ambient gas in the heavy elements \citep{Searle:1972p328}. This model also
makes the approximations that the stellar lifetimes and timescale for complete
mixing of nucleosynthetic products are negligible in comparison to the timescale
on which the metallicity evolves (``instantaneous recycling''). The simple model
makes a specific prediction regarding the metallicity and system properties: 

\begin{equation}
Z = y \ln (M_{total}/M_{gas}),
\end{equation}

\noindent in this equation $Z$ is the metal abundance, $y$ is the fraction
of the stellar mass converted to heavy elements (the 
{\em yield}), and $M_{total}$ = $M_{gas}$ + $M_{stars}$. 

Although this model is most appropriate for the low-mass galaxies, 
it can also be applied to large disk galaxies if
concentric radii are treated as independent zones. However, it does not explain
the observed gradients, so modifications such as radial flows, matter exchange
with an outside reservoir (infall and outflow), or a variable stellar initial
mass function, have been proposed as modifications to the model
\citep{Matteucci:1989p2141,Dopita:1990p2143}.

The relative abundances of nitrogen and oxygen are of particular interest, since
they are synthesized in different astrophysical sites. Oxygen is synthesized in
massive stars and distributed into the interstellar medium by Type II
supernovae, while the origin of nitrogen is more problematic. A distinction is
frequently made between ``primary'' nucleosynthetic products, which can be
synthesized directly from H and He in Population III stars, and ``secondary''
products, which require a ``seed'' heavy nucleus to be initially present in the
star where its synthesis occurs. By this definition, oxygen is a primary
species. Nitrogen is secondary when made as a by-product of CNO-cycle hydrogen
burning. According to the simple closed-box model, the abundance of a secondary
species is quadratic, so that if N is secondary and O primary, then (N/H) $\propto$
(O/H)$^2$, or (N/O) $\propto$ (O/H). The N/O ratio does appear to approach this behavior,
for \hh regions with moderately high O/H values in M\,101
\citep{TorresPeimbert:1989p2005} and in the Milky Way. However,
below a certain values of O/H, it appears that N/O is constant; these
low-metallicity \hh regions occur mostly in low-mass galaxies. 
Thus, it is becoming clear that nitrogen is not purely a secondary
nucleosynthetic product. Indeed, N may be produced within intermediate-mass
stars by an effectively primary process, if C synthesized within the star by the
triple-alpha reaction is later subjected to the CN cycle. 
Nitrogen made by this process would be primary, but there might be a time-delay 
in building up its abundance relative to the nuclear products of supernovae, 
because of the longer lifetimes of the source stars.

The other elements measured in extragalactic \hh regions, S, Ne, and Ar, are
not likely to be dominated by secondary processes. They might still, however,
vary differently than oxygen, if they were produced in stars of different mass
ranges and the IFM varied or the timescales for enrichment
differed substantially. There are known variations in the abundance ratios of
certain elements. For example the fact that the iron-group is deficient relative
to oxygen in Population II stars is thought to reflect an origin for the former
chiefly in Type I supernovae, which originate in long-lived progenitors, as
opposed to synthesis of oxygen in massive stars and Type II supernovae.

In the context of chemical evolution models, \citet{Garnett:2002p339}
studied the metallicity-luminosity and metallicity-rotation speed correlations 
for spiral and irregular galaxies 
for a sample of spiral and irregular galaxies having well-measured
abundance profiles, distances, and rotation speeds. 
He finds that the O/H-V$_{rot}$ relation shows a change in slope at a rotation speed
of about 125 km\,s$^{-1}$. At faster V$_{rot}$, there appears to be no relation 
between average metallicity and rotation speed. At
lower V$_{rot}$, the metallicity correlates with rotation speed. This change in
behavior could be the result of increasing loss of metals from the smaller
galaxies in supernova-driven winds. The idea was tested by looking at the
variation in effective yield, derived from observed abundances and gas fractions
assuming closed box chemical evolution. The effective yields derived for spiral
and irregular galaxies increase by a factor of 10-20 from V$_{rot}$ $\sim$ 5
to 300 km\,s$^{-1}$, asymptotically increasing to approximately constant $y_{eff}$
for V$_{rot}$ $\sim$ 150 km\,s$^{-1}$. 
The trend suggests that galaxies with V$_{rot}$ $\sim$ 100-150 km\,s$^{-1}$
may lose a large fraction of their supernova ejecta, while galaxies above this
value tend to retain metals. The determination of effective yields as function
of galactic radius and its interpretation stands as one of the main studies
in order to discriminate among different physical effects which may affect the
chemical evolution of a galaxy.

\section{Goals of this dissertation}
\label{sec:goals}

As described in this chapter, the study of chemical abundances has undergone a
remarkable development in the last decades thanks mostly to important
observational efforts that have focused on the derivation of physical and
chemical properties of emission line \hh regions in galaxies by spectroscopic
techniques. The main motivation common to all of these observations is to use
the chemical information as one of the means at our disposal to understand the
physical processes at play in the formation and evolution of galaxies in the
universe.

Hitherto, most spectroscopic studies in nearby objects have been limited by
the number of objects sampled, the number of \hh regions observed and the
coverage of these regions within the galaxy surface. In order to increase
significantly the number of \hh regions sampled in any given galaxy we require
the combination of high quality multi-wavelength data and wide field
spectroscopy. The advent of multi-object and integral field spectrometers with
large field of view now offer us the opportunity to undertake a new generation
of observations, based on samples of scores to hundreds of \hh regions and
full 2-dimensional (2D) coverage. These sort of data would enable to test,
confirm and extent the previous body of results from small sample studies,
while at the same time open up a new frontier of studying the 2D
metallicity structure of disks and the intrinsic dispersion in metallicity, or
to test and strengthen the diagnostic methods that are used to measure the \hh
region abundances in galaxies, among other issues.

The scientific core of this dissertation is based on an observational project
conceived to tackle the problem of the 2D spectroscopic coverage of
the whole galaxy surface. New techniques in imaging spectroscopy (or integral
field spectroscopy, IFS) provide a powerful tool for studying the
small and intermediate scale-size variation in line emission and stellar
continuum in nearby well-resolved galaxies. 
We designed a project to take advantage of these new observational
techniques in order to assemble a unique spectroscopic sample from which we
could study, with unprecedented detail, the star formation and gas chemistry
across the surface of a galaxy.
The observations consist of Integral Field Unit (IFU)
2D spectroscopic mosaics of a representative sample of nearby
galaxies ($D\,<\,100$ Mpc) with a projected angular size of less than 10 arcmin. 
The mosaics were constructed using the unique instrumental capabilities of the
Postdam Multi Aperture Spectrograph, PMAS
\citep{Roth:2005p2463} in the PPAK mode
\citep{Verheijen:2004p2481,Kelz:2006p338} at the German-Hispanic Astronomical
Centre at Calar Alto (CAHA), Spain. The PMAS fibre PAcK (PPAK) is one of the
world's widest integral field unit with a field-of-view (FOV) of
74\,$\times$\,65 arcseconds that provides a semi-contiguous regular sampling
of extended astronomical objects.
This project represents the first attempt to obtain 2D spectra of
the whole surface of a galaxy in the nearby universe. The spectroscopic
mosaicing comprises more than 50\,000 spectra in the optical wavelength range.

This observational project was devised as a scientific international consortium, 
the members are world-leading experts in their respective fields, including
star formation and chemical abundances of galaxies, active galactic nuclei,
multiwavelength observations of emission line regions and 2D
spectroscopy. The project was entitled: the {\bf P}PAK {\bf I}FS {\bf N}earby
{\bf G}alaxies {\bf S}urvey, or {\bf PINGS}. The P.I. of this project is
Prof. Robert C. Kennicutt Jr. at the Institute of Astronomy, University of
Cambridge. The collaborators of the consortium are: Dr. Ángeles Díaz at the
Universidad Autónoma de Madrid, Spain; Dr. Anna Pasquali at the Max-Plank
Institut für Astronomie in Heidelberg, Germany; Dr. Sebastián S. Sánchez at
CAHA, Spain; Benjamin Johnson and Caina Hao at the University of Cambridge,
UK.

The primary scientific objectives of this dissertation are to use the PINGS
observations to obtain pixel-resolved emission-line maps across the disks of
the galaxies to study the 2D abundance distribution and on
characterising the relations between these abundance properties and the
physical properties of the parent galaxies. By targeting virtually every \hh
region in the galaxies, as a consequence of
the nearly complete spatial coverage of the IFUs, we are able to test for the
first time the systematic dependences of the
strong-line abundances on the size, luminosity, surface brightness, and other
properties of the \hh regions. In that respect, the PINGS observations and the
subsequent analysis represent a leading leap in the study of the chemical
abundances and the global properties of galaxies, 
information which is most relevant for interpreting observations at all
redshift sources accessible with the current technology.

\subsection{Structure of the dissertation}

The structure of this thesis is as follows:
In \S\,2, I discuss the importance of Integral Field Spectroscopy (IFS) in
astrophysics, including an explanation of the technique with their advantages
and pitfalls, a description of the available instrumentation in the world by
the time this project was envisaged, and the selection criteria of the
telescope-instrument chosen for this project. I also include a brief review to
previous works that have attempted to obtain 3D chemical abundance information
in galaxies.
In \S\,3, I present the sample of galaxies and the selection criteria followed
according to the scientific objectives established for this dissertation. This
chapter also includes a full description of the logistics and observations,
explaining the telescope set-up and the particular observing technique adopted
for this project.
In \S\,4, I explain the reduction process of the IFS raw data, the additional
corrections implemented in this project and the improvements with respect to
previous pipelines, particularly regarding the flux calibration and the
sky-subtraction. The possible sources of errors and uncertainties are
addressed, together with an explanation of the techniques implemented to
minimise them.
In \S\,5, I present the integrated spectra of the PINGS sample, obtained by
co-adding the spectra from their corresponding mosaics. Comparisons with
previously published data are included. An analysis of the ionized gas
component is performed, together with the techniques and methodologies
implemented in order to derive the physical parameters of the integrated
gas-phase of the galaxies sample.
In \S\,6, I present a complete 2D spectroscopic study for a selected number of
galaxies from the sample. Selected  \hh regions previously
observed are compared with spectra extracted from the PINGS sample. A set of
emission line maps calculated from each galaxy is presented, including a
quantitative description of the 2D distribution of the physical properties
inferred from them. Then, a detailed, spatially-resolved spectroscopic
analysis of the selected galaxies is performed, based on different spectral
samples extracted from the full IFS mosaics of the galaxies. 
Several diagnostic diagrams and the state-of-the-art abundance diagnostic
techniques are used to obtain the 2D distribution of the physical properties
and chemical abundances of the selected sample.
Finally, in \S\,7 I present the general conclusions of this dissertation,
including some planned paths of future investigation.











\vspace*{1cm}

\chapter*{Thesis content}

\vspace*{1cm}

\begin{center}

\Large
The whole thesis is not included in astro-ph due to file size limitations.\\
\vspace*{1cm}

The full contents can be found at:\\

\vspace*{0.5cm}

\url{http://www.dspace.cam.ac.uk/handle/1810/224843}

\end{center}



\bibliographystyle{mn2e}

\bibliography{astroph}

\end{document}